\documentclass[twocolumn]{article}
\usepackage{amsmath}
\usepackage{subcaption}
\usepackage{graphicx} 
\usepackage{authblk}
\usepackage{booktabs}
\usepackage{float}
\usepackage{appendix}
\usepackage{xcolor}
\def\dd{\text{d}}
\def\e{\text{e}}

\begin{document}
\title{Gravitational lensing of the wormhole in the Eddington-inspired Born-Infeld spacetime with a cosmic string}
\author[1]{Xin-Fei Li \thanks{xfli@gxust.edu.cn}}
\author[2]{Lei-Hua Liu \thanks{liuleihua8899@hotmail.com}}
\author[1]{Yan-Zhi Meng}
\author[1]{Shu-Qing Zhong}
\author[1]{Li-Juan Zhou}
\affil[1]{School of Sciences, Guangxi University of Science and Technology, 545026 Liuzhou, Guangxi, China}
\affil[2]{Department of Physics, College of Physics, Mechanical and Electrical Engineering,
Jishou University, 416000 Jishou, China}

\date{}
\maketitle

\begin{abstract}
In this work we study gravitational lensing  of the wormhole in the Eddington-inspired Born-Infeld (EiBI) spacetime that incorporates with a cosmic string. It was found that
the presence of cosmic string can enhance the light deflection in strong field limit, compared to the case of the Eills-Bronnikov wormhole. The magnification effects of this composite structure could cause some substantial impacts on the angle separation between the first and the rest of the images, and their relative brightness. Furthermore, based on these observables, we model some observable aspects in the strong and the weak field limits. The presence of a cosmic string can affect some distinguishable observables compared to the wormhole without cosmic string. This work could deepen our understanding of the spacetime structure of the wormhole in EiBI spacetime with one-dimensional topological defects.
\end{abstract}
\section{Introduction}
General Relativity (GR) has been remarkably successful in describing a wide range of gravitational phenomena. However, it exhibits certain limitations that may motivate the exploration of alternative theories of gravity. One of the most pressing issues within GR is the singularity problem, which implies that freely falling objects inevitably encounter singularities, such as those at the centers of black holes. Moreover, GR does not inherently account for the observed acceleration of the universe or the singularity associated with the Big Bang, prompting the need for more comprehensive gravitational theories.

In this context, modified gravity theories, such as Eddington-inspired Born-Infeld (EiBI) gravity, offer promising alternatives. EiBI gravity is inspired by the Eddington gravitational action and Born-Infeld nonlinear electrodynamics and asymptotically approaches GR in the low-energy regime \cite{Deser:1998rj, Banados:2010ix} (see \cite{BeltranJimenez:2017doy} for a review). Notably, within the Palatini framework \cite{Olmo:2011uz}, EiBI gravity can yield various singularity-free black hole and WH solutions without the need for exotic matter or quantum effects \cite{Olmo:2013gqa, Lobo:2013adx, Wei:2014dka, Shaikh:2018yku, Olmo:2020fnk}. Also, refs. \cite{Yang:2013hsa, Du:2014jka} have analyzed  perturbational stability of EiBI gravity possible observations beyond $\Lambda$CDM models in cosmology. These characters allow  for the construction of traversable wormholes (WHs), which are hypothetical structures that facilitate rapid interstellar travel between distant regions of spacetime \cite{Morris:1988cz, Einstein:1935tc, Fuller:1962zza, Bronnikov:1973fh}.

In the context of EiBI gravity, WH solutions can exist in various topological defects, such as monopoles, cosmic strings, and domain walls \cite{Avelino:2020irs}. Cosmic strings would have formed during phase transitions in the early universe \cite{Kibble:1976sj} or as a result of brane collisions in D-brane inflationary scenario \cite{Sarangi:2002yt, Copeland:2003bj, Dvali:2003zj}. The tension of cosmic strings (energy per unit length) is determined by the energy scale of symmetry breaking, which may play a significant role in seeding the large-scale structure of the universe \cite{Vilenkin:1981iu}. Recent research has indicated that gravitational waves generated by cosmic strings could provide a plausible cosmological origin for the stochastic background of gravitational waves detected by pulsar timing arrays \cite{Chang:2019mza, Sousa:2020sxs, NANOGrav:2023gor, EPTA:2023sfo, Reardon:2023gzh, Xu:2023wog}. It is worth mentioning that in other contexts, WHs associated with cosmic strings have been studied serve as a spacetime connectors between two WHs \cite{Clement:1995ya}. The WHs incorporated with a cosmic string are able to maintain stability at certain perturbations \cite{Eiroa:2004at, Eiroa:2016rdd, Eiroa:2019lhh, Bejarano:2006uj}. The presence of a black hole or a WH pierced by a cosmic string can significantly alter the structure of spacetime \cite{Bambhaniya:2024hzb}.

Gravitational lensing is a powerful observational tool that can probe the spacetime structure of wormholes and other massive objects \cite{Bartelmann:1999yn, Treu:2010uj,Vagnozzi:2022moj}. This phenomenon occurs when light rays emitted by distant stars are deflected by the gravitational fields of massive objects as gravitational lenses. In the weak limit, gravitational lensing manifests as slight distortions in light paths, while strong lensing refers to light rays being bent several times around massive objects before reaching an observer. Initial analytical studies on strong lensing were conducted by Bozza for Schwarzschild black hole \cite{Bozza:2001xd}. Then, Tsukamoto introduced an improved method to study light deflection in a general asymptotically flat, static, spherically symmetric spacetime \cite{Tsukamoto:2016qro, Tsukamoto:2016jzh}. Note that recently a method to express deflection angle independent of the coordinates was suggested in \cite{Igata:2025taz}. Thereafter, gravitational lensing has been applied to detect properties of varied massive objects in several contexts involving black holes \cite{Eiroa:2005ag, Zhao:2016kft, Canfora:2018isz, Molla:2022izk, Kuang:2022xjp, Liu:2022lfb, Gao:2022cds, Gao:2023ipv, Gao:2023sla, Molla:2023hou, QiQi:2023nex, Zhang:2023rsy, Duan:2023gvm, Junior:2024vdk}, naked singularities \cite{Virbhadra:2002ju, Virbhadra:2007kw}, WHs \cite{Nandi:2006ds, Ovgun:2018fnk, Shaikh:2019jfr, Tsukamoto:2021apr, Zhou:2022dze, Godani:2021aub, Cai:2023ite, Hsieh:2024eph,Jimenez:2024npb}, modified gravity \cite{Cheng:2021hoc, Atamurotov:2022slw, Tsukamoto:2021caq, Ali:2021psk, Kumar:2021cyl, Hensh:2021nsv, Ghosh:2022mka, Junior:2023xgl, Chen:2023trn}, regular black holes \cite{Kumaran:2023brp, dePaula:2023ozi, Zhang:2023oui, Olmo:2023lil, Xie:2024dpi} and galaxies \cite{Liao:2022gde, Virbhadra:2022ybp, Toscani:2023gdf}.

Recently, some studies have been focused on exploring in both weak and strong lensing effects for WHs with topological defects \cite{Jusufi:2018waj, Furtado:2020puz, Soares:2023err}. The influence of string tension on light deflection in black holes pierced by cosmic strings has been examined in various contexts \cite{Wei:2011bm, Jusufi:2017vew, Bambhaniya:2024hzb}. The latest works have shown that cosmic string could increase the deflection of light in the spacetime of WH in EiBI gravity \cite{Ahmed:2023rqd, Ahmed:2023dvc, Ahmed:2023qnn}. While a comprehensive study on the deflection of light ray in the strong limit for the WH in EiBI gravity with cosmic string remains unexplored. This could reveal valuable insights on the lensing images and provide potential observational aspects.

Thus, in this work, we will perform a general study of the WH in EiBI gravity in the presence of a cosmic string, which  could also describe the global monopole (GM) whose spacetime structure corresponds to a solid angle deficit. The deflection of light will be addressed not only in the weak field limit, but also in the strong field limit for the first time, to our knowledge. Additionally, we will calculate several observables  and discuss the potential for their detection. It should also be noted that the results can contribute to a deeper understanding of the interplay between the spacetime structure influenced by cosmic strings and WHs in the context of EiBI gravity.

The outlines of this work are arranged as follows: In Sect. 2, the spacetime of the EiBI gravity with a cosmic string is introduced, whose lensing effects will be briefly reviewed. Then, the magnification of lensing image will be studied. In Sect. 3, The light deflection is studied in the strong field limit. Some observables are considered, such as the magnification of images, the angular separation, and the brightness ratio among the different images. In Sect. 4, by modeling the WH with a cosmic string in our galaxy center, some observable quantities are predicted in  strong and weak field limit. The conclusion is drawn in the last section.

\section{Lensing of EiBI gravity with a cosmic string}
\subsection{Spacetime of EiBI gravity with a cosmic string}
Firstly, we will briefly review the simplest traversable WH in EiBI gravity. Following that, the spacetime of WH with a static cosmic string will be considered.

The general line element of static WH solution in background of the EiBI gravity in the spherical coordinates $(t, r, \theta, \varphi)$ is characterized by
\begin{equation}
    \dd s^2=-\e ^{2\Phi(r)}\dd t^2+\frac{\dd r^2}{1+\frac{B(r)}{r}}+r^2(\dd\theta^2+\sin^2 \theta \dd\varphi^2), \label{whcs1}
\end{equation}
where $\Phi(r)$ is the redshift function for an infalling observer and $B(r)$ the shape function. For the simplest WH, we consider the redshift function to be zero and that, in order to prevent the presence of an event horizon, the shape function $B(r_0)=\varepsilon$, where $\varepsilon$ is constant in the throat of the WH $r_0$. If $\varepsilon=0$, the space-time of eq. \eqref{whcs1} characterizes a flat spacetime. For the case $\varepsilon>0$, \eqref{whcs1} corresponds to a GM; for $\varepsilon<0$, it corresponds to the Morris-Thorn WH \cite{Morris:1988cz}.

The spacetime of cosmic string in the spherical coordinates is given by
\begin{eqnarray}
    \dd s^2=-\dd t^2+\dd r^2+r^2\left(\dd \theta^2+a^2\sin^2 \theta\dd\varphi^2 \right), \label{flatspacetime}
\end{eqnarray}
where the dimensionless string parameter $a$ is the cosmic string parameter. Note that $a=1-4 G \mu$, where $G$ is the Newton constant and $\mu$ the energy density per unit length. String tension $G\mu$ is typically very small in the natural unit. As $a\to 1$, the spacetime reduces to Minkowski spacetime  \cite{Linet:1985mvm, Vilenkin:2000jqa}.

Furthermore, one can introduce the spacetime of the WH with a cosmic string sitting on the center of the WH or the GM by redefining the azimuthal angle in such a way that $\varphi\to \varphi'=a\varphi$. Therefore, by this transformation, the metric \eqref{whcs1} is written as \cite{Ahmed:2023dvc},
\begin{equation}
\dd s^{2}=-\dd t^{2}+\left(1+\frac{\varepsilon}{r^{2}}\right)^{-1}\dd r^{2}+r^{2}(\dd \theta^{2}+a^{2}\sin^{2}\theta \dd \varphi^{2}), \label{whcs2}
\end{equation}
which is known as the conical Morris-Thorn WH space-time of cosmic string. This EiBI gravity background space-time with a cosmic string is asymptotically flat since $\varepsilon/r \to 0$ as $r\to \infty$.

The deficit angle caused by cosmic string makes some considerable impacts on the structure on the spacetime and the behavior of null geodesics. Therefore, this work aims to study how cosmic string affects the deflection of light rays and to provide some potential observational aspects.

\subsection{Angle deflection for the WH with a cosmic string}
\begin{figure}[H]
    \centering
    \includegraphics[width=0.9\linewidth]{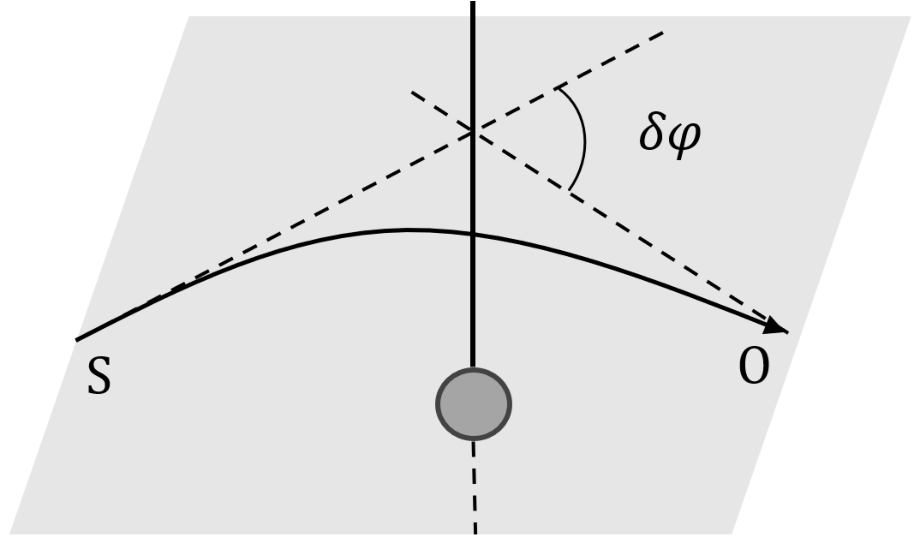}
    \caption{Scheme of the deflection of light $\delta \varphi$ for the WH with a cosmic string perpendicular on the equilateral plane. Cosmic string and  WH are represented by the black straight line and  gray disk, respectively.}
    \label{fig:deflection}
\end{figure}

In this section, we will briefly review the null geodesic of the WH in EiBI gravity with a cosmic string, as presented in ref. \cite{Ahmed:2023dvc}. Photons emitted from source S are bent when passing near the lens where a straight cosmic string pierces the WH, as shown in Fig. \ref{fig:deflection}.

\begin{figure}[H]
    \centering
    \includegraphics[width=0.9\linewidth]{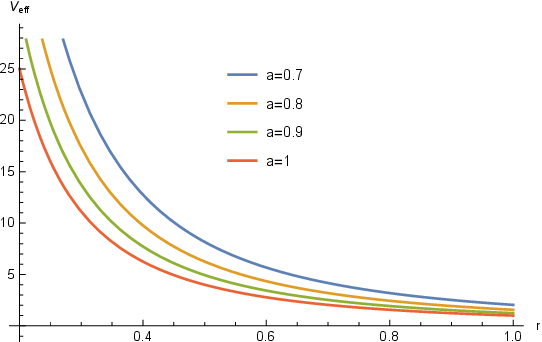}
    \caption{Effective potential vs radius for various cosmic string parameters. Here choosing $L=1$.}
    \label{fig:potential}
\end{figure}

To obtain the null geodesic associated with \eqref{whcs2}, the variational method will be adopted. The length $\mathcal{S}$ of a smooth curve on a spacetime with metric \eqref{whcs2} is given by
\begin{equation}
    \mathcal{S}=\int \dd x\mathcal{L}=\int \dd\tau \sqrt{g_{\mu\nu}\frac{\dd x^{\mu}}{\dd \tau}\frac{\dd x^{\mu}}{\dd \tau}},
\end{equation}
where $\tau$ is the affine parameter of the curve. According to eq. \eqref{whcs2}, for $\theta=\frac{\pi}{2}$, the Lagrangian of a photon in the presence of the WH with a cosmic string is
\begin{align}
\mathcal{L}&=-\dot{t}^{2}+\left(1+\frac{\varepsilon}{r^{2}}\right)^{-1}\dot{r}^{2}+a^{2}r^{2}\dot{\varphi}^{2}. \label{Lagrangian}
\end{align}
The corresponding Euler-Lagrange equation for the coordinates $t$ and $\varphi$ results in the two conserved quantities
\begin{equation}
   L\equiv a^{2}r^{2}\frac{\dd \varphi}{\dd \tau},\ \  E\equiv\frac{\dd t}{\dd \tau}. \label{EnerMoment}
\end{equation}
where $E$ is the energy  parameter and $L$ the angular momentum.
For the null geodesic, it is well known that $\mathcal{L}=0$. Then, the \eqref{Lagrangian} leads to
\begin{align}
\left(\frac{\dd r}{\dd \tau}\right)^{2}&=\left(1+\frac{\varepsilon}{r^{2}}\right)\left(E^{2}-\frac{L^{2}}{r^{2}a^{2}}\right). \label{geodesic}
\end{align}
This equation can be interpreted as describing the one-dimensional motion of a particle with energy $E$ subject to an effective potential $V_{eff}=\frac{L^{2}}{r^{2}a^{2}}$. In Fig. \ref{fig:potential}, we plot the effective potential for some values of $a$, which implies that the potential increases as the string parameter decreases.

The turning point satisfies $\frac{\dd r}{\dd \tau}=0$, which leads to
\begin{equation}
    r_0=\frac{L}{aE}=\frac{b}{a}, \label{turingpoint}
\end{equation}
where the impact parameter $b\equiv\frac{L}{E}$. Note that the turning point is larger than the case without a cosmic string, where $a=1$. This can modify the deflection of the trajectory of the light ray. It is worth mentioning that in the case of the WH with a cosmic string, where $\varepsilon<0$, the solution of eq. \eqref{geodesic} for $r=r_0$ has a minimal radius given by $r=\sqrt{|\varepsilon|}$; therefore, photons carrying sufficient energy could pass to the other side of the WH.

By substituting eq. \eqref{EnerMoment} and \eqref{turingpoint} into eq. \eqref{geodesic}, we have
\begin{equation}
    \frac{\dd \varphi}{\dd r}=\frac{b}{a^{2}r^{2}}\cdot\frac{1}{\sqrt{(1+\frac{\varepsilon}{r^{2}})(1-\frac{b^{2}}{r^{2}a^{2}})}}. \label{dphidr}
\end{equation}
By symmetry, the contribution to $\delta \varphi$ before and after the turning point is equal. Hence, by integration from the turning point to the infinity, eq. \eqref{dphidr} becomes
\begin{equation}
    \delta \varphi=2\int^\infty_{r_0}\frac{b}{a^{2}}\cdot\frac{\dd r}{\sqrt{(r^2+\varepsilon)(r^2-\frac{b^{2}}{a^{2}})}}. \label{deltaphi1}
\end{equation}

After integrating eq. \eqref{deltaphi1}, one obtains
\begin{equation}
    \delta\varphi = \frac{2}{a}K\left(\frac{-a^2\varepsilon}{b^{2}}\right), \label{deltavarphi}
\end{equation}
where $K(x)$ is the complete first kind elliptical integral.  When $x\ll 1$ or $x\to0$, the complete first kind elliptical integral is expanded as
\begin{align}
    K(x)=\frac{\pi}{2}\sum^{\infty}_{n=0}\left(\frac{(2n!)}{2^{2n}(n!)^2}\right)^2x^n.
\end{align}

The deflection angle \eqref{deltavarphi} is valid for the WH with a cosmic string case $\varepsilon>0$ and the GM with a cosmic string case $\varepsilon<0$. For the GM case $x<0$, $K(x)$ is an incomplete elliptical integral of the first type, given in terms of the parameter itself rather than the modulus. In the GM with a cosmic string case, one could write $x=\frac{-a^2\varepsilon}{b^{2}}=-y$, with $y=\frac{a^2\varepsilon}{b^{2}}>0$. We are left with
\begin{equation}
    K(x)=K(-y)=\frac{1}{\sqrt{y+1}}K\left(\frac{y}{y+1}\right).
\end{equation}

\begin{figure}[H]
    \centering
    \includegraphics[width=0.9\linewidth]{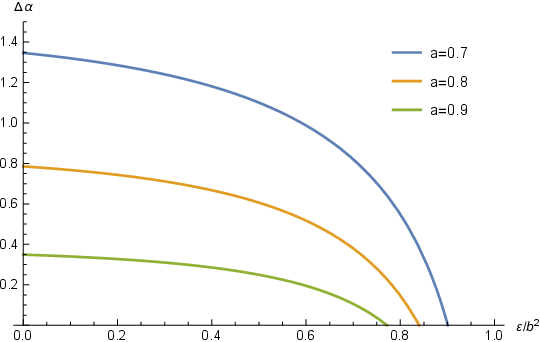}
    \caption{Differences of deflection angel $\Delta\alpha$ for various $a$ vs. the dimensionless parameter $\varepsilon/b^2$. }
    \label{fig:AngleDeference}
\end{figure}

The deflection angel is given by $\alpha=\delta\varphi-\pi$,
which is expressed alternatively as following:
    \begin{eqnarray}
    \alpha=\begin{cases}
    \frac{2}{a}K\left(-a^2\varepsilon/b^2\right)-\pi,\ \mbox{($\varepsilon<0$)},\\
    \frac{2}{a\sqrt{a^2\varepsilon/b^2+1}}K\bigg(\frac{a^2\varepsilon/b^2}{a^2\varepsilon/b^2+1}\bigg)-\pi,\ \mbox{($\varepsilon>0$)}. \label{deflectionangle}
    \end{cases}
    \end{eqnarray}
 As illustrated in Fig. \ref{fig:AngleDeference}, the deflection angle increases as the string parameter increases; while the angular difference, denoted as $\Delta \alpha \equiv \alpha(a \neq 1) - \alpha(a = 1)$, exhibits a decreasing trend with increasing values of $\alpha$ and $\varepsilon/b^2$.

For the case of WH $(\varepsilon<0)$ with a cosmic string in the weak field limit where $b\ll |\varepsilon|$,
the deflection angle can be expanded as
\begin{eqnarray}
   \alpha= \left(\frac{1}{a}-1\right)\pi-\frac{a}{4}\frac{\varepsilon}{b^{2}}\pi+\mathcal{O}\left(\frac{\varepsilon}{b^2}\right)^2.
\end{eqnarray}

\section{Light deflection in the strong field limit}

As discussed in the previous section, eq. \eqref{deltavarphi} is valid for the WH with a cosmic string and the GM  with a cosmic string. The light deflection keep finite for the GM case
($\varepsilon>0$), while the deflection angle diverges for the WH case when $\frac{a^2\varepsilon}{b^2}\to 1$ since the elliptical integral $K(x)$ diverges as $x\to1$.
This limit  is called the strong field, which corresponds to the situation that the turning point towards the approximation of the radii of the WH throat. Hence, from now on we will concentrate on the deflection angle and the observational quantities related to them in the strong field limit.

\begin{figure}
    \centering
    \includegraphics[width=1\linewidth]{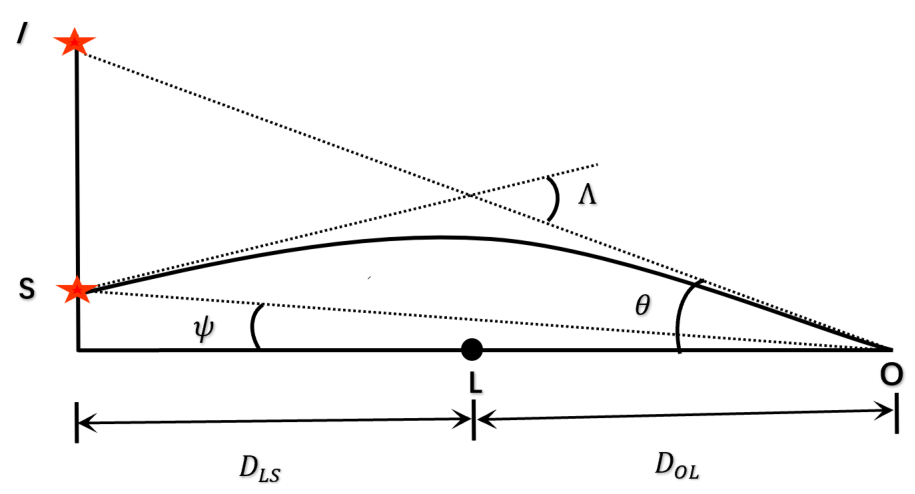}
    \caption{Visual lensing profile. The light emitted from the source S is bend due to the WH associated with cosmic string located at L, and then passing towards to the observer O. I is the image of S observed from O. $\Lambda$ is the deflection angle. $D_{LS}$ denotes the distance between the worm hole and the source, and $D_{OL}$ denotes the distance between the WH and the observer.}
    \label{fig:lensing}
\end{figure}

For simplicity, we choose $\varepsilon=-\lambda^2$, where the parameter $\lambda$ denotes the radius of the WH throat. From eq. \eqref{deltavarphi}, one arrives at
\begin{equation}
    \delta \varphi =\frac{2}{a}K\left(\frac{a^2\lambda^2}{b^2}\right)-\pi.
\end{equation}

In the strong field limit, \textit{i.e.,} $a\lambda \to\ b$,  the deflection angle of light  diverges logarithmically \cite{Nascimento:2020ime}, which becomes
\begin{align}
 \delta\varphi=-\frac{1}{a}\log\left(\frac{b}{a\lambda}-1\right)+\frac{3}{a}\log2-\pi \notag \\
 +\mathcal{O}\left(1-\frac{a\lambda}{b}\right)\log\left(1-\frac{a\lambda}{b}\right).   \label{deltavarphi2}
 \end{align}
As shown in Fig. \ref{fig:lensing}, following  ref. \cite{Virbhadra:1999nm}, we obtain the following relationship
\begin{equation}
   \tan\psi=\tan\theta-\frac{D_{LS}}{D_{OS}}\left[\tan\theta+\tan(\Lambda-\theta)\right],
\end{equation}
where $D_{OS}=D_{OL}+D_{LS}$ is the distance between the observer and the source.

Now let's assume that the lens L and the source S in Fig. \ref{fig:lensing} are almost perfectly aligned. Even though the angular positions of the source and the images are small, the light ray may have circled around the WH several times before reaching the observer. Thus,  $\Lambda$ is approximate to a multiple of $2\pi$ \cite{Bozza:2001xd}, \textit{i.e.}, $\Lambda=2\pi n+\Delta \Lambda_n$, where $\Delta \Lambda_n$ is the deflection angle after circling $n$ times around the lens, which leads to $\tan(\Lambda-\theta)\sim \Delta \Lambda_n-\theta$. With these information, the lensing equation in the strong field limit is written as
\begin{equation}
\psi=\theta-\frac{D_{LS}}{D_{OS}}\Delta\Lambda_{n}.    \label{psi}
\end{equation}
When $\theta$ is very small, the critical impact parameter in cosmic string spacetime,
\begin{equation}
  b=D_{LS}\sin(a\theta)\simeq aD_{LS}\theta. \label{impactparameter}
\end{equation}
Note that the critical impact parameter was modified by the cosmic string; its derivation refers to Appendix A. Then,
the angular deflection eq. \eqref{deltavarphi2} is re-expressed as
\begin{equation}
 \Lambda(\theta)=-\frac{1}{a}\log\left(\frac{D_{OL}\theta}{\lambda}-1\right)+\frac{3}{a}\log2-\pi. \label{Lambda}
 \end{equation}

In order to get $\Delta \Lambda_n$ in the lens equation \eqref{psi}, we should expand $\Lambda(\theta)$ around $\theta^0_n$, where $\Lambda(\theta^0_n)=2n\pi$, \textit{i.e.},
\begin{equation}
 \Lambda(\theta)=\Lambda(\theta_{n}^{0})+\frac{\partial\Lambda}{\partial\theta}\Big|_{\theta=\theta_{n}^{0}}(\theta-\theta_{n}^{0}),
\end{equation}
 \begin{equation}
\Delta\Lambda_{n}\equiv\frac{\partial\Lambda}{\partial\theta}\Big|_{\theta=\theta_{n}^{0}}(\theta-\theta_{n}^{0}).    \label{DeltaAlphan}
\end{equation}
Using $\Lambda(\theta^0_n)=2\pi n$ in eq. \eqref{Lambda}, one finds
\begin{equation}
  \theta_{n}^{0}=\frac{\lambda}{D_{OL}}\left(1+\frac{8}{\e^{(2n+1)a\pi}}\right).  \label{theta-n0}
\end{equation}
Then substituting eq. \eqref{Lambda} and eq. \eqref{theta-n0} into eq. \eqref{DeltaAlphan}, we obtain

\begin{equation}
 \Delta\Lambda_{n}=-\frac{\e^{(2n+1)a\pi}D_{OL}}{8a\lambda}\Delta\theta_{n}.   \label{DetalLambda-n}
\end{equation}
After substituting eq. \eqref{DetalLambda-n} into the lens equation \eqref{psi}, one yields to
\begin{equation}
 \theta_{n}\simeq\theta_{n}^{0}+\frac{8a\lambda D_{OS}(\psi-\theta_{n}^{0})}{D_{LS}D_{OL}\e^{(2n+1)a\pi}}.  \label{theta-n}
\end{equation}
From the equation above, as $a \to 1$, the angular position of the $n$th image is recover to Ellis-Bronnikov WH \cite{Tsukamoto:2016qro}, corresponding to be no cosmic string; as $a <1$, the angular positions of the relativistic images are modified by the string parameter,  which are larger than the case of the Ellis-Bronnikov BH where $a=1$.

Now let's consider the magnification of images. The total flux from the $n$th lensed image is proportional to the magnification of the image, which is defined by
\begin{equation}
    \mu_{n}=\bigg|\frac{\psi}{\theta}\frac{\dd \psi}{\dd \theta}\bigg|_{\theta_{n}^{0}}^{-1}. \label{magnification-n}
\end{equation}
Substituting eq. \eqref{DetalLambda-n} and eq.  \eqref{theta-n} into eq. \eqref{magnification-n}, we have
\begin{equation}
 \mu_{n}=\frac{8a}{\psi}\frac{D_{OS}}{D_{LS}}\left(\frac{\lambda}{D_{OL}}\right)^{2}\frac{1}{\e^{(2n+1)a\pi}}\left(1+\frac{8}{\e^{(2n+1)a\pi}}\right),
\end{equation}
\begin{figure}[H]
    \centering
    \includegraphics[width=0.9\linewidth]{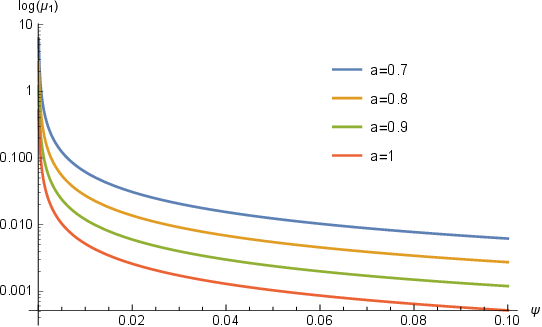}
    \includegraphics[width=0.9\linewidth]{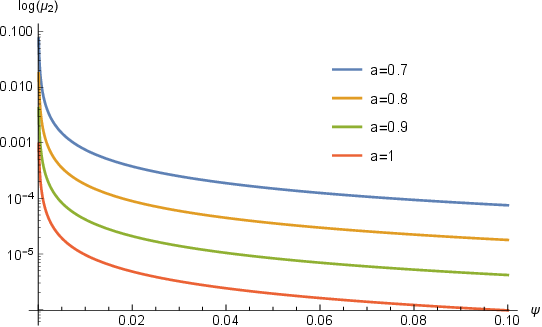}
    \caption{Magnification of the first and the second image in terms of $\psi$. Here $D_{OL}=D_{LS}=5 Mpc$, $D_{OS}=10 Mpc$ and $\lambda=1 Mpc$.}
    \label{fig:StrongMagnf1}
\end{figure}
The expression of $\mu_n$ above reveals that the magnification of $n$-th image increases exponentially with increasing $n$.  It is clear that the first image, denoted as $\mu_1$, has a significantly higher magnification compared to the others, for example, as one can see the lower sub-figure $\mu_2$ shown in  Fig. \ref{fig:StrongMagnf1}. Fig. \ref{fig:StrongMagnf1} also illustrates that the magnification of the images decreases as the string parameter $a$ increases.  Furthermore, as $\psi $ approaches zero-indicating a stronger alignment between the source and the lens-the magnification of the images increases significantly. The magnification of the first image is nearly 100 orders of magnitude greater than that of the second image for the same string parameter.

Eq. \eqref{theta-n} and \eqref{magnification-n} respectively denote the positions of the relativistic images and the magnifications in terms of the WH throat radii $\lambda$ and the cosmic string parameter $a$. Inversely, it is important to introduce some observational methods, which are closely tied to the parameters characterizing the WH with a cosmic string. Such methods not only provide a promising approach to search for WHs but also serve as valuable tools to detect the theory beyond GR. In particular, the observables proposed by Bozaa could help to distinguish between a GR WH and an EiBI WH \cite{Bozza:2001xd}.  Bozza has introduced the following observables:
\begin{equation}
   s=\theta_{1}-\theta_{\infty} \ \text{and}\ \ R=\frac{\mu_{1}}{\sum_{n=2}^{\infty}\mu_{n}} ,
\end{equation}
where $s$ is the angular separation between the first and the rest of the relativistic images, and $R$ the brightness ratio between  the flux of the first image and the flux of the rest ones. From eq. \eqref{theta-n}, we have
\begin{equation}
s\simeq \frac{8\theta_{\infty}}{\e^{3a\pi}}    \label{eqs}
\end{equation}
and from eq. \eqref{magnification-n}, we can show that
\begin{equation}
  R=\frac{\e^{-3a\pi}+8\e^{-6a\pi}}{\e^{-5a\pi}+8\e^{-10a\pi}}.
\end{equation}
Note that both $s$ and $R$ are independent of the wormhole throat but the cosmic string parameter $a$. When $a\to1$,
we come to
\begin{equation}
    R\simeq \e^{2a\pi}. \label{eqR}
\end{equation}
As seen from eq. \eqref{eqs} and eq. \eqref{eqR}, the presence of the cosmic string increases the angular separation $s$, while it decreases the ratio $R$ since $a<1$.

\section{Observational aspects from WH with a cosmic string}
Aiming to explore potential observational aspects of the EiBI WH with a cosmic string,  we will consider three astrophysical scenarios: one for the strong field limit and two for the weak field limit. The ranges of the WH radii $\lambda$ and the cosmic string parameter $a$ should be estimated. Initially, based on the time delay between the signals of GW170817 and GRB170817A in a background
Friedmann-Robertson-Walker universe, an upper bound was imposed on the Eddington parameter $(\varepsilon)$, which is given by $|\varepsilon|\leq10^{37}\rm m^2$ \cite{Jana:2017ost}. The radius of the WH throat in this context is given by $\lambda =\sqrt{|\varepsilon|}$, leading to an estimation of $\lambda \leq 10^{15}\rm km$. Moreover, the width of cosmic string could be negligible compared to the radii of WH throat. According to the particle models producing cosmic strings, the string tension is estimated by $G\mu=10^{-6}(\frac{\eta}{10^{16}~\rm GeV})^2$. Assuming that cosmic strings were formed due to the symmetry breaking at the grand unified theory energy scale with $\eta\sim10^{16}~\rm GeV$, we find $a=1-4\times 10^{-6}$.

\subsection{ Strong field limit}
In the strong field limit, we will model the WH in the presence of a cosmic string using data from Sagittarius A*, located at the center of our galaxy. The distance between Sagittarius A* and Earth is approximately $D_{OL}=8.5~\rm Kpc$, and the mass of Sagittarius A* is $4.4\times 10^6~M_{\odot}$. In Table 1, we present the values of the observables, $\theta_\infty$, $s$ and $\Tilde{R}$ for various values of the WH throat radius $\lambda$. The observable $s$ is obtained from \eqref{eqs} and $\Tilde{R}=2.5\text{log}_{10}R$, where $R$ is given by eq. \eqref{eqR}. This redefinition is convenient for comparing our results with those found in the literature.
    \begin{table}[H]
    \centering
    	\label{t0}
    	\begin{tabular}{c c l}
    		\hline
    		$\lambda(km)$ & $\theta_{\infty}$(micro-arcsecs)             & s(micro-arcsecs)  \\  \bottomrule
    		$10^9$        & $0.78642$    & $0.000508$              \\
    		$10^{10} $   & $7.8642$  &  $0.00508$               \\
    		$10^{11}$    & $78.642$   & $0.0508$                \\
    		$10^{12}$ & $786.42$   & $0.508$  \\ \hline
    	\end{tabular}
     \caption{Einstein Radii/angle for Bulge  Lensings}
    \end{table}
    For the Schwarzschild black hole, the observables for the scenario applied here have been computed by \cite{Virbhadra:1999nm, Furtado:2020puz}, yielding values of $\theta_\infty=26.547~$ micro-arcsecs, $s=0.03322$ and $\Tilde{R}=6.822$ magnitudes. For the WH radii in the range of $10^{10}km<\lambda<10^{11}km$, we found that the magnitude of the critical angle and angular separation for the Schwarzschild black hole case and the EiBI WH with a cosmic string are of the same order. Our results indicate that both the angular separation $\theta_{\infty}$ and the strength $s$ are larger as the radii of WH increases for $10^9~\rm km<\lambda<10^{12}~\rm km$. Regarding the magnification, $\Tilde{R}=6.822$, it is independent of the throat radius and is comparable to the values for the Schwarzschild black hole case. Hence, the angular separation could be a feature to distinguish the WH with a cosmic string and the Schwarzschild black hole case  in the strong field limit region \cite{Soares:2023err}.

    Future high-resolution experiments (such as the next-generation Event Horizon Telescope (ngEHT) and LISA) will significantly enhance the precision of measurements of strong-field gravitational lensing effects (such as angular separation \( s \)).  Based on the model presented in this paper and the future experiments, some discussions on the strong lensing case are presented as follows:
\begin{itemize}
 \item Sensitivity to Angular Separation $s$: Eq. \eqref{eqs} indicates that \( s \propto \e ^{-3a\pi} \), which increases significantly with the decrease of the cosmic string parameter \( a \) (i.e., as \( G\mu \) increases). If future observations can detect non-Schwarzschild deviations of \( s \) at higher resolutions (such as sub-microarcsecond levels) (Table 1), they can directly rule out the extreme case of \( a \to 1 \) (i.e., \( G\mu \to 0 \)), thereby providing an upper limit constraint on \( G\mu \). Such as ngEHT with angular resolution \( \sim 5 \, \text{micro-arcsecs} \) \cite{Johnson:2023ynn}, it could resolve separations \( s \geq 0.1 \, \mu\text{micro-arcsecs} \), corresponding to \( G\mu < 10^{-8} \) for \( \lambda \sim 10^{11} \, \text{km} \) (via Eq. \eqref{eqs}).

 \item Multi-messenger Observations: Recent pulsar timing arrays have begun to explore constraints on the cosmic string gravitational wave background ($G\mu <5.1\times 10^{-10}$) \cite{Wu:2021kmd}, and $G\mu \in [1.43,15.3]\times 10^{-12}$ with a small reconnection probability for supertrings \cite{Wu:2023hsa}. The Advanced LIGO and Virgo O3 dataset could set bound on  $G\mu <10^{-15}$  in nano-hertz band \cite{LIGOScientific:2021nrg}. Combining the lensing effects presented in this paper, future joint analyses of electromagnetic and gravitational wave signals can cross-verify the physical range of \( G\mu \)  (such as ruling out models with $G\mu>10^{-7}$).
\end{itemize}
\subsection{Weak field limit}
\begin{figure}
    \centering
    \includegraphics[width=0.9\linewidth]{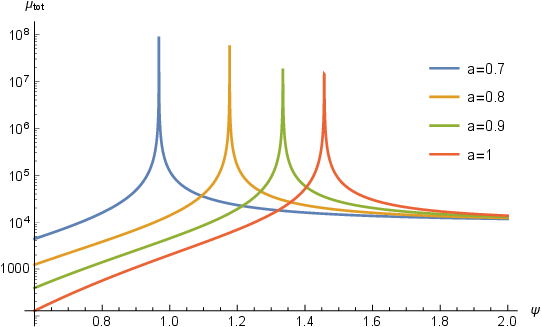}
    \caption{Plot for magnification of image in terms of $\psi$ with string parameters $a=0.7, 0.8, 0.9, 1$. Without loss of generality,  we fix $D_{OL}=D_{LS}=5 Mpc$, $D_{OS}=10 Mpc$ and $\lambda=0.01 Mpc$.}
    \label{fig:WeakMagnf}
\end{figure}

In the weak field limit, where  $\frac{a^2\lambda^2}{b^2} \ll 1$, light ray passing through the WH will be slightly bent. The deflection angle is given by eq. \eqref{deltavarphi}, which yields
\begin{equation}
    \Lambda\sim(\frac{1}{a}-1)\pi+\frac{a}{4}\frac{\lambda^{2}}{b^{2}}\pi. \label{weakangle2}
\end{equation}
In our context, we consider the WH case is considered where $\varepsilon=-\lambda^2$ in eq. \eqref{weakangle2}.  The angular position $\theta$ can be obtained from eq. \eqref{psi}, eq. \eqref{impactparameter} and eq. \eqref{weakangle2}, leading to
\begin{equation}
     \theta^{3}-\left(\psi+\frac{\pi D_{LS}}{D_{OS}}\frac{1-a}{a}\right)\theta^{2}-\frac{\pi}{4}\frac{\lambda^2}{aD_{OL}D_{LS}}=0.  \label{thetaeq}
\end{equation}
According to the total magnification of lensed image $\mu_\text{tot}=\sum_i\left|\frac{\psi}{\theta_i}\frac{\dd\psi}{\dd \theta_i}\right|^{-1}$, where $\theta_i$ are solutions of eq. \eqref{thetaeq}. There is only one real solution for eq. \eqref{thetaeq}, which corresponds to one image. The numerical  magnification of image in terms of $\psi$ is shown in Fig. \ref{fig:WeakMagnf}.  The magnification of the image in the weak field limit is found to decrease as the string parameter $a$ increases.

Particularly, considering observer, the WH with a cosmic string and the source are perfectly aligned \textit{i.e.,} $\psi=0$, and $a\to1$. The solution of the equation above is given by
\begin{align}
    \theta_{E}&=\left(\frac{\pi\lambda^{2}}{4aD_{OL}D_{LS}}\right)^{1/3}\notag \\&+(1-a)\left[\left(\frac{\pi\lambda^{2}}{4aD_{OL}D_{LS}}\right)^{1/3}+\frac{\pi D_{LS}}{3D_{OS}}\right].
\end{align}
The Einstein radius $R_E$ is obtained from \eqref{impactparameter}, which is
\begin{equation}
    R_E=aD_{OL}\theta_E.
\end{equation}
\begin{table}[H]
\centering
	\label{t1}
	\begin{tabular}{c c l}
		\hline
		$\lambda(km)$      & $R_E(km)$                &  $\theta_E(mas)$ \\ \bottomrule
	      $10^9$       & $7.83\times10^{11}$      &  $1.31 \times 10^3$              \\
		$10^{10} $   & $2.46\times 10^{12}$     &  $4.10\times10^{3}$               \\
		$10^{11}$    & $1.02\times 10^{13}$     &  $1.71\times10^{4}$                \\
		$10^{12}$    & $4.63\times 10^{13}$     &  $7.73\times 10^{4} $  \\ \hline
	\end{tabular}
 \caption{Einstein Radii/angle for Bulge  Lensing}
\end{table}

\begin{table}[H]
\centering
	\label{t2}
	\begin{tabular}{c c l}
		\hline
		$\lambda(km)$                 & $R_E(km)$              & $\theta_E(mas)$ \\    \bottomrule
		$6\times10^9$           & $4.81\times10^{12}$    & $1.29\times 10^2$            \\
		$10^{10}$               & $595\times 10^{12}$    & $1.59\times 10^{2}$          \\
		$10^{11}$               & $2.02\times 10^{13}$   & $5.41\times 10^{3}$          \\
		$10^{12}$               & $8.66\times 10^{13}$   & $ 23.17\times 10^{3} $      \\ \hline
	\end{tabular}
 \caption{Einstein Radii/angle for LMC Lensing}
\end{table}
Now let us estimate the observables, the Einstein radii $R_E$ and the Einstein angle  $\theta_E$ via taking some reasonable values for the model parameters. Following the example of \cite{Abe:2010ap}, we consider the lensing effects of a bulge star and the Large Magellanic Cloud (LMC). For a bulge star, we adopt the following parameters: $D_{OS}=8~\text{kpc}$ and $D_{OL}=4~\text{kpc}$; for the LMC:  $D_{OS}=50~\text{kpc}$ and $D_{OL}=25~\text{kpc}$. Based on these parameters, some values of the observable for two different scenarios are presented in Table 2 and Table 3, respectively. It is verified that the predicted results in EiBI with a cosmic string are detectable  when the radius of the throat of the WH $a$ being the order of $10^9~\text{km}$ or more.
In the equatorial plane, the Einstein radii and angle separation of EiBI with a cosmic string are both larger than the EiBI WH without cosmic string in the weak limit, which is of the order of $100~\text{arcsec}$ and  within the observable range. Hence, the WH with a cosmic string can be distinguished from the WH without cosmic string by the Einstein radii and angle separation.

\section{Conclusion }
In this work, we analytically and numerically study the deflection angle of photons of WH within Eddington-inspired Born-Infeld  (EiBI) gravity in the presence of a cosmic string. After a brief review of the deflection of light caused by the WH and the global monopole with a cosmic string,  the deflection angle for both cases is substantially subject to the critical impact parameter. Then, we concentrate on the strong field limit, where the results show that the angular separations between the first images and the rest of the images increase as cosmic string parameter increases, while the relative brightness ratio $R$ decreases; Both of these quantities are independent of the WH throat.

For our further investigation, the WH with a cosmic string was modeled by choosing some realistic parameters from the observational data of Sagittarius A*. Our results indicates that in the strong field limit, the WH with a cosmic string can be distinguished from the typical Schwarzschild black hole. Furthermore, the string parameter enhances the angular separation and the flux strength of the first image. In the weak field limit, there is only one real image. And the magnification of the image in the weak field limit is found to decrease as the string parameter $a$ increases. By combining the data of a bulge star and the Large Magellanic Cloud, the Einstein radii and angular separation of EiBI with a cosmic string are enhanced in the equatorial plane. These observables provide potential capability to distinguish the EiBI WH with a cosmic string from the one without string, deepening our understanding the spacetime structure of the EiBI gravity with one-dimensional topological defects.

Advances in future observational techniques (such as ngEHT, SKA and LISA) will provide stricter tests for this model. For example, sub-microarcsecond resolution can directly measure deviations in angle separation and limit the upper bound of string tension; while multi-messenger observations (such as gravitational waves + lensing) hold the promise of revealing unique characteristics of the interaction between cosmic strings and EiBI gravity.

\begin{appendices}
\section{Deriving $b=D_{LS}\sin(a\theta)$}
For the WH with a cosmic string in \eqref{whcs2}, as $r\to \infty$, it becomes the spacetime of  cosmic string:
\begin{equation}
\dd s^{2}=-\dd t^{2}+\dd r^{2}+r^{2}(\dd \theta^{2}+a^{2}\sin^{2}\theta \dd \varphi^{2}), \label{flatcs2}
\end{equation}
The trajectory of a photon on the equilateral plane is
\begin{equation}
\left(\frac{\dd u}{\dd \varphi}\right)^2+a^2u^2-\frac{a^2}{b^2}=0, \label{cstrajectory}
\end{equation}
where $u\equiv\frac{1}{r}$. Then, differentiate on \eqref{cstrajectory} with respect on $u$, one comes to
\begin{equation}
    \frac{\dd ^2u}{\dd \varphi^2}+a^2u=0.
\end{equation}
The general solution above is
\begin{equation}
    u=C_1 \sin{a\varphi}+C_2 \cos{a \varphi}.
\end{equation}
By imposing the conditions, \textit{i.e.}, $u(\varphi=0)=0$ and $u(\varphi=\frac{\pi}{2})=\frac{1}{b}$, the constants $C_1$ and $C_2$ are determined,  which is $C_1=\frac{1}{b\sin(\frac{\pi}{2}a)}$ and $C_2=0$.  $a$ is closed to 1, leading to $C_1\simeq \frac{1}{b}$. we have
\begin{equation}
    u=\frac{1}{b}\sin{a \varphi}. \label{u}
\end{equation}
By choosing $u=\frac{1}{D_{LS}}$, we are left with
\begin{equation}
    b=D_{LS}\sin(a\theta).
\end{equation}
\end{appendices}
\section{Acknowledgements}
X.-F. Li is supported by Youth Program of Natural Science Foundation of Guangxi with Grant No. 2021GXNSFBA075049 and Doctor Start-up Foundation of Guangxi University of Science and Technology with Grant No. 19Z21. L.-H. Liu is supported by National Natural Science Foundation of China Grant No. 12165009, Hunan Natural Provincial Science Foundation Grant NO. 2023JJ30487. S.-Q. Zhong is supported by the starting Foundation of Guangxi University of Science and Technology Grant No. 24Z17. L.-J. Zhou is supported by Natural Science Foundation of Guangxi with Grant  No. 2025GXNSFAA069552 and National Natural Science Foundation of China Grant No. 11865005.

\end{document}